\documentclass[aps, prl, floatfix, reprint]{revtex4-1}

\usepackage{graphicx}
\usepackage{amsmath}
\usepackage{color}
\usepackage{amssymb}
\usepackage{textcomp}
\usepackage{bm}
\usepackage{hyperref}

\hypersetup{
	colorlinks = true, 
	urlcolor   = blue, 
	linkcolor  = blue, 
	citecolor  = blue, 
}

\begin{document}
	\title{Magnetic Bloch Skyrmion Transport by Electric Fields in a Composite Bilayer}
	\author{Zidong Wang}
	\email{Zidong.Wang@auckland.ac.nz}
	\author{Malcolm J. Grimson}
	\email{m.grimson@auckland.ac.nz}
	\affiliation{Department of Physics, University of Auckland, Private Bag 92019, Auckland, New Zealand}
	\date{\today}
	
	\begin{abstract}
		We investigate a mechanical method to manipulate magnetic Bloch Skyrmions by applying an electric field in a composite chiral-magnetic (CM)/ferroelectric (FE) bilayer. The magnetoelectric coupling at the interface allows the electric field to stimulate magnetic ordering. Therefore it offers the possibility to generate Skyrmions [Phys. Rev. B \textbf{94}, 014311 (2016)]. Here, we design a movable and localized electric field source to drive Skyrmion transport along the bilayer. A traveling velocity of the electric field source must be carefully chosen to show the stability and efficiency of this process. The effects of high speed operation will be discussed.
	\end{abstract}
	
	\maketitle

\textit{Introduction.} - Magnetic Bloch Skyrmions have been introduced theoretically \cite{N.442.797,NN.8.899} and observed experimentally \cite{S.323.915,NP.7.713} in many works. They occurs in materials which the magnetic order breaks the centrosymmetric nanostructure due to the existence of asymmetric exchange interaction (well-known as the Dzyaloshinskii-Moriya interaction). Magnetic Skyrmions have been detected experimentally from the range of ${\text{MnSi}}$ \cite{S.323.915}, ${\text{FeGe}}$ \cite{NM.10.106}, ${\text{MnGe}}$ \cite{NN.8.723}, ${\text{GaV}}_{4}{\text{S}}_{8}$ \cite{NM.14.1116},  ${\text{Cu}}_{2}{\text{OSeO}}_{3}$ \cite{PRL.108.237204}, ${\text{Fe-Co-Si}}$ alloys \cite{PRB.81.041203} and ${\text{Mn-Fe-Ge}}$ alloys \cite{NN.8.723}. They offer a great potential for applications in spintronic memory devices, due to their self-protection behavior. To use magnetic Skyrmions as information holders, the current high interest is to control their motion. Several non-mechanical methods have been investigated, such as by using electric current dynamics \cite{NM.15.501}, spin-polarized currents \cite{NN.8.152,NN.8.742}, magnetic fields \cite{PRB.92.020403}, temperature gradients \cite{PRL.111.067203,PRL.112.187203}, and magnons \cite{PRB.90.094423}. But a mechanical method, however, has not been explored. In this paper, a mechanical technique to move magnetic Bloch Skyrmions collinearly with a mobile electric field source is investigated.

\begin{figure}
	\includegraphics[width = 250px, trim = 60 600 30 600, clip]{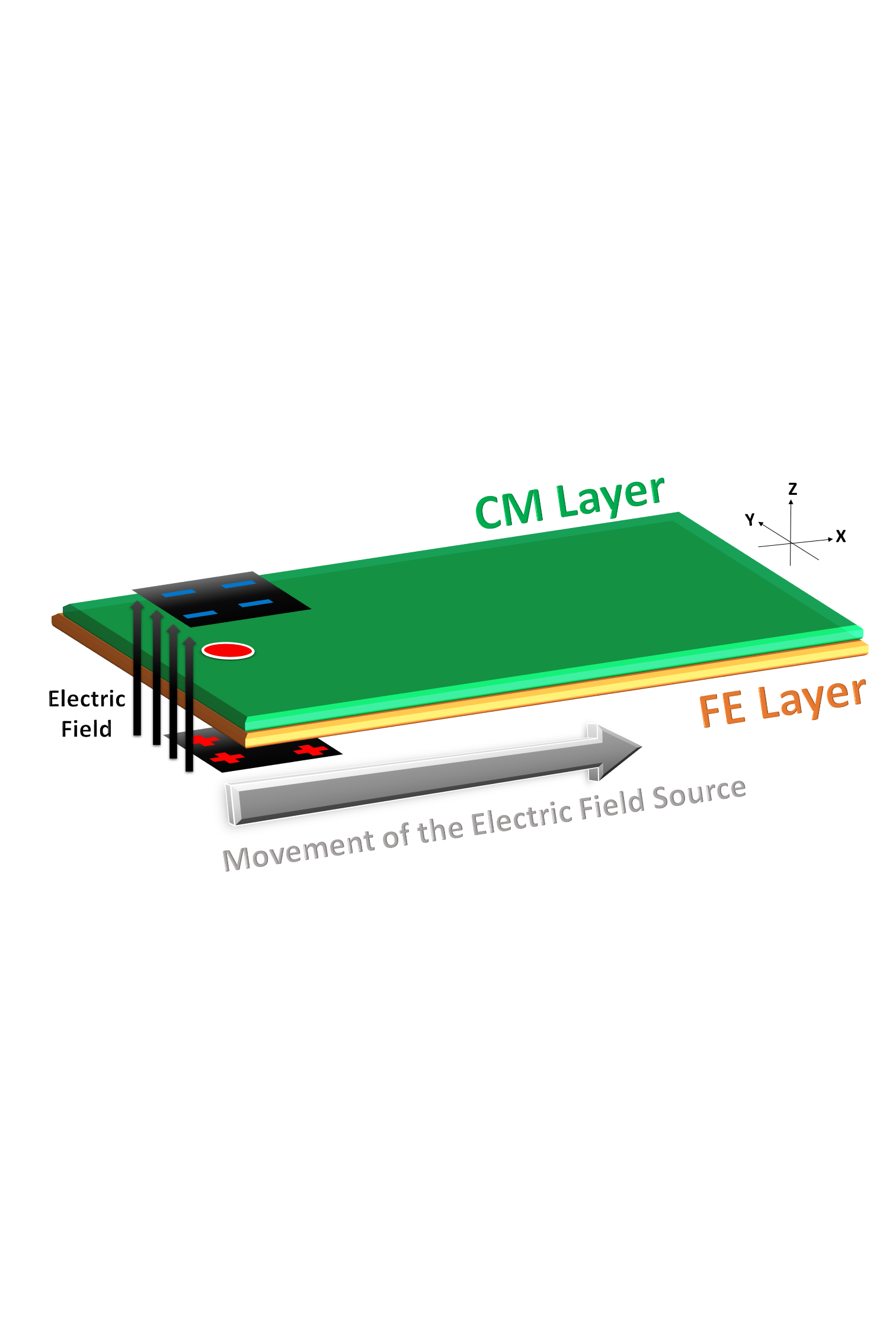}
	\caption{ A schematic illustration of a composite bilayer consisting of a CM layer and a FE layer stacked at the interface. A localized electric field source can be moved longitudinally along the bilayer. }
	\label{Fig.1}
\end{figure}

A previous study has discussed magnetic Bloch Skyrmions induced by the electric field in a composite bilayer in Ref.~\cite{PRB.94.014311}. Here, we pursue a microscopic approach to Skyrmion transport by moving the electric field source in a plane parallel to the bilayer. In \textbf{FIG.~\ref{Fig.1}}, we (i) construct a bilayer lattice model, which contains a chiral-magnetic (CM) layer with classical magnetic spins and a ferroelectric (FE) layer with electric pseudospins, both of layers are glued by a strong magnetoelectric (ME) coupling, (ii) attach a localized electric field source, which can travel longitudinally along the bilayer film, (iii) carry out a spin dynamics method to determine the time-response behaviors of the magnetic spins and the electric pseudospins, respectively. \textbf{Movie 1} in the Supplemental Material shows an animation of the dynamics \cite{Supplementary.8}. The results show the creation and propagation of magnetic Bloch Skyrmions, and we discuss the stability and efficiency of the traveling velocity to Skyrmions. This mechanical technique provides a guide for designing and developing a Skyrmion transport channel in future spintronic devices.

\textit{Model and Simulation.} - The total Hamiltonian is constructed by effective Hamiltonians in the CM and FE layers and the interfacial interaction, as: $\mathcal{H} = \mathcal{H}_{\text{CM}} + \mathcal{H}_{\text{FE}} + \mathcal{H}_{\text{ME}}$.
Firstly, the Hamiltonian in CM layer $\mathcal{H}_{\text{CM}}$ is described by a classical Heisenberg model, as
\begin{eqnarray}
	\mathcal{H}_{\text{CM}} & = & -J_{\text{CM}} \sum_{i,j} [ \bm{S}_{i,j} \cdot ( \bm{S}_{i+1,j} + \bm{S}_{i,j+1}) ]\nonumber\\
	&   & -D_{\text{CM}} \sum_{i,j} [ \bm{S}_{i,j} \times \bm{S}_{i+1,j} \cdot \hat{x} + \bm{S}_{i,j} \times \bm{S}_{i,j+1} \cdot \hat{y} ]\nonumber\\
	&   & -K_{\text{CM}} \sum_{i,j} (S_{i,j}^z)^2,
	\label{Eq.CM}
\end{eqnarray}
where $\bm{S}_{i,j} = (S_{i,j}^x , S_{i,j}^y , S_{i,j}^z)$ represents the local magnetic spin is used to characterize the magnetic moment, which is a normalized vector, i.e., $\| \bm{S}_{i,j} \| = 1$, and $i,j \in [1,2,3,...,N]$ characterizes the location of each spin in the CM layer. The first term shows the symmetric exchange interaction between neighbor spins, where $J_{\text{CM}}^* = J_{\text{CM}} / k_B T$ represents the dimensionless magnetic exchange coupling. The second term shows the asymmetric exchange interaction, and $D_{\text{CM}}^* = D_{\text{CM}} / k_B T$ represents the dimensionless Dzyaloshinskii-Moriya coefficient, with $\hat{x}$ and $\hat{y}$ are the unit vectors of the \textit{x} and \textit{y} axes, respectively. Weak magnetocrystalline anisotropy exists in CM materials \cite{PRL.113.107203}. In general, the last term shows a perpendicular magnetic anisotropy, with $K_{\text{CM}}^* = K_{\text{CM}} / k_B T$ representing the dimensionless uniaxial anisotropy coefficient along the \textit{z} axis.

In the FE layer, we employ a pseudospin model to solve the dynamical behaviors of electric polarization \cite{JAP.118.124109}. The local electric moment is replaced by the pseudospin, shown as $\bm{P}_{k,l} = (P_{k,l}^x , P_{k,l}^y , P_{k,l}^z)$, which is regarded as a continuous vector, and $k,l \in [1,2,3,...,N]$ characterizes each pseudospin’s location. The distinction between a pseudospin and a classical spin, is the variable size and no precession of the pseudospin. Since the electric polarization is defined as the dipole moment density in dielectric materials. The dipole moment density $p$ is proportional to the external electric field, as $p = \epsilon_0 \chi_e E_{\text{ext}}$ \cite{C.Kittel}. In the pseudospin model, the size of each electric pseudospin is proportional to the magnitude of its effective field, as $\| \bm{E}_{k,l}^{\text{eff}} = \delta \mathcal{H} / \delta \bm{P}_{k,l} \|$ \cite{JAP.119.124105}. Hence, $\| \bm{P}_{k,l} \| = \epsilon_0 \Xi_e \| \bm{E}_{k,l}^{\text{eff}} \|$, where $\Xi_e$ is the dimensionless pseudo-scalar susceptibility. Consequently, the electric pseudospin has a variable size as does the behavior of electric dipole. The Hamiltonian thus be described by a transverse Ising model \cite{SSC.1.132}, as
\begin{eqnarray}
\mathcal{H}_{\text{FE}} & = & -J_{\text{FE}} \sum_{k,l} [ P_{k,l}^z ( {P}_{k+1,l}^z + {P}_{k,l+1}^z ) ]\nonumber\\
&   & -\Omega_{\text{FE}} \sum_{k,l} (P_{k,l}^x)\nonumber\\
&   & -\epsilon_0 \chi_e E_{\text{ext}}^z \sum_{\tilde{k},\tilde{l}} P_{\tilde{k},\tilde{l}}^z,
\label{Eq.FE}
\end{eqnarray}
where $J_{\text{FE}}^* = J_{\text{FE}} / k_B T$ represents the dimensionless electric exchange coupling along the Ising \textit{z} direction. $\Omega_{\text{FE}}^* = \Omega_{\text{FE}} / k_B T$ represents the dimensionless transverse field along the +\textit{x} axis, which is a in-plane field and perpendicular to the Ising \textit{z} direction. $E_{\text{ext}}^* = \epsilon_0 \chi_e E_{\text{ext}}^z / k_B T$ represents the dimensionless electric field, which been applied in the +\textit{z} direction, $\epsilon_0$ is the electric permittivity of free space, and $\chi_e$ is the dielectric susceptibility. A particular site of locations $\tilde{k} , \tilde{l}$ presents the pseudospins which in the presence of electric field. Remember that applied electric field is mobile, and spatially attached to the bulk of bilayer film to reduce the edge effects (i.e., edge-merons) in simulations \cite{JAP.120.203903}.

Interfacial effects between CM and FE layers are caused by a ME coupling. The mechanism behind it can be understood by a strain-stress effect. Since in ferroelectrics, a mechanical strain internally generated from the applied electric field due to the reverse piezoelectric effect. It physically exerts on the CM layer, resulting in a magnetization due to the inverse magnetostrictive effect. Consequently, a series of electric-mechanical-mechanical-magnetic effects constitute the converse ME effect, which emphasizes the influence of electric polarization on the magnetization at interface. Nan \textit{et al.} have given a detailed study of this behavior \cite{JAP.103.031101}. The analytic expression of ME effect can be linear or nonlinear, particularly with respect to the thermal effect \cite{PRB.92.134424}. In this paper, we only account for low-energy excitations between the CM and FE layers and so we restrict ourselves to the linear expression of ME interaction, as
\begin{eqnarray}
	\mathcal{H}_{\text{ME}} = -g_{\text{ME}} \sum_{(i,j)(k,l)} (S_{i,j}^z P_{k,l}^z),
\end{eqnarray}
where $g_{\text{ME}}^* = g_{\text{ME}} / k_B T$ represents the dimensionless strength of ME coupling. This was discussed by Spaldin \cite{PRB.93.195167}. The ME coupling strength is, however, currently unknown.

The dynamics of magnetic spins has been studied by the well-known Landau-Lifshitz-Gilbert equation \cite{JPD.48.305001}, which numerically solves the rotation of a magnetic spin in response to its torques,
\begin{equation}
	\frac{\partial \bm{S}_{i,j}}{\partial t} = -\gamma_{\text{CM}} [\bm{S}_{i,j} \times \bm{H}_{i,j}^{\text{eff}}] - \lambda_{\text{CM}} [\bm{S}_{i,j} \times (\bm{S}_{i,j} \times \bm{H}_{i,j}^{\text{eff}})],
	\label{Eq.LL}
\end{equation}
where $\gamma_{\text{CM}}$ is the gyromagnetic ratio, and $\lambda_{\text{CM}}$ is the phenomenological damping term of CM materials. $\bm{H}_{i,j}^{\text{eff}} = - \delta \mathcal{H} / \delta \bm{S}_{i,j}$ is the magnetic effective field acting on each spin, which is the functional derivative of the total Hamiltonian with respect to each magnetic spin.

In the FE layer, pseudospins are used to describe the location of electric dipoles. The electric dipole moment is a measure of the separation of positive and negative charges along the Ising \textit{z} direction. It is scalar. Therefore, only the \textit{z} component of pseudospin represents the real polarization, and the time evolution of pseudospins is expected to perform a precession free trajectory \cite{JAP.118.124109}, as
\begin{equation}
	\frac{\partial \bm{P}_{k,l}}{\partial t} = - \lambda_{\text{FE}} [\bm{P}_{k,l} \times (\bm{P}_{k,l} \times \bm{E}_{k,l}^{\text{eff}})],
	\label{Eq.Pseudo-LL}
\end{equation}
where $\lambda_{\text{FE}}$ is the phenomenological damping term in the FE structure. $\bm{E}_{k,l}^{\text{eff}} = - \delta \mathcal{H} / \delta \bm{P}_{k,l}$ is the electric effective field acting on each pseudospin. \textbf{Movie 2} in the Supplemental Material shows animations for comparing the trajectories of magnetic spin and electric pseudospin \cite{Supplementary.8}.

\textit{Results.} - Dimensionless parameters are used for the numerical simulations: $J_{\text{CM}}^* = 1$, $D_{\text{CM}}^* = 1$, $K_{\text{CM}}^* = 0.1$, $J_{\text{FE}}^* = 0.8$, $\Omega_{\text{FE}}^* = 0.1$, $g_{\text{ME}}^* = 0.4$, $\gamma_{\text{CM}}^* = 1$, and $\lambda_{\text{CM}}^* = \lambda_{\text{FE}}^* = 0.1$. A number of $N = 30 \times 90$ magnetic spins/electric pseudospins in each layer and free boundary conditions are used. Landau-Lifshitz-Gilbert equations are solved by a fourth-order Range-Kutta method.
A marginal electric field is applied to order the FE and CM domain walls before the dynamics, then we apply the localized electric field with a magnitude $E_{\text{ext}}^* = 10$, perpendicular to the surface. Electric pseudospins quickly complete realignment, but the response of magnetic spins has a delay. The generation process of a Skyrmion in the bulk of CM layer is summarized in \textbf{FIG.~\ref{Fig.2}} (\textbf{Movie 3} in the Supplemental Material \cite{Supplementary.8}).
Subsequently, this field source is moved along the bilayer with a traveling velocity. The velocity is measured as $v^* = \Delta N / \Delta t^*$, where $\Delta N$ corresponds to spatial movement to equivalent locations (i.e., spin-site), and $\Delta t^*$ is a dimensionless time step. \textbf{Figure~\ref{Fig.3}} shows a series of diagrams about the Skyrmion transport in the CM layer collinearly following the polarized pseudospins in the FE layer, for a traveling velocity of $v^* = 0.02$. \textbf{Movie 3} in the Supplemental Material presents the full dynamical process \cite{Supplementary.8}. In this propagation process, we can see the Skyrmion track deflecting to the bottom edge is due to the Skyrmion Hall effect \cite{NP.2016}. The behavior of a Skyrmion is topologically like a spinning disk, it generates a Magnus force when traveling longitudinally. So it induces a transverse force during the translational motion of the Skyrmion. \textbf{Figure~\ref{Fig.3}} furthermore shows the electric polarization reflecting the passage of field source. But the magnetization has a component that is non-collinear with the electric response, and shows a spin spiral alignment, due to the existence of a finite Dzyaloshinskii-Moriya interaction. CM crystals have a non-centrosymmetric structure enables the magnetic ordering to be broken.

\begin{figure}
	\includegraphics[width = 250px, trim = 60 510 0 0, clip]{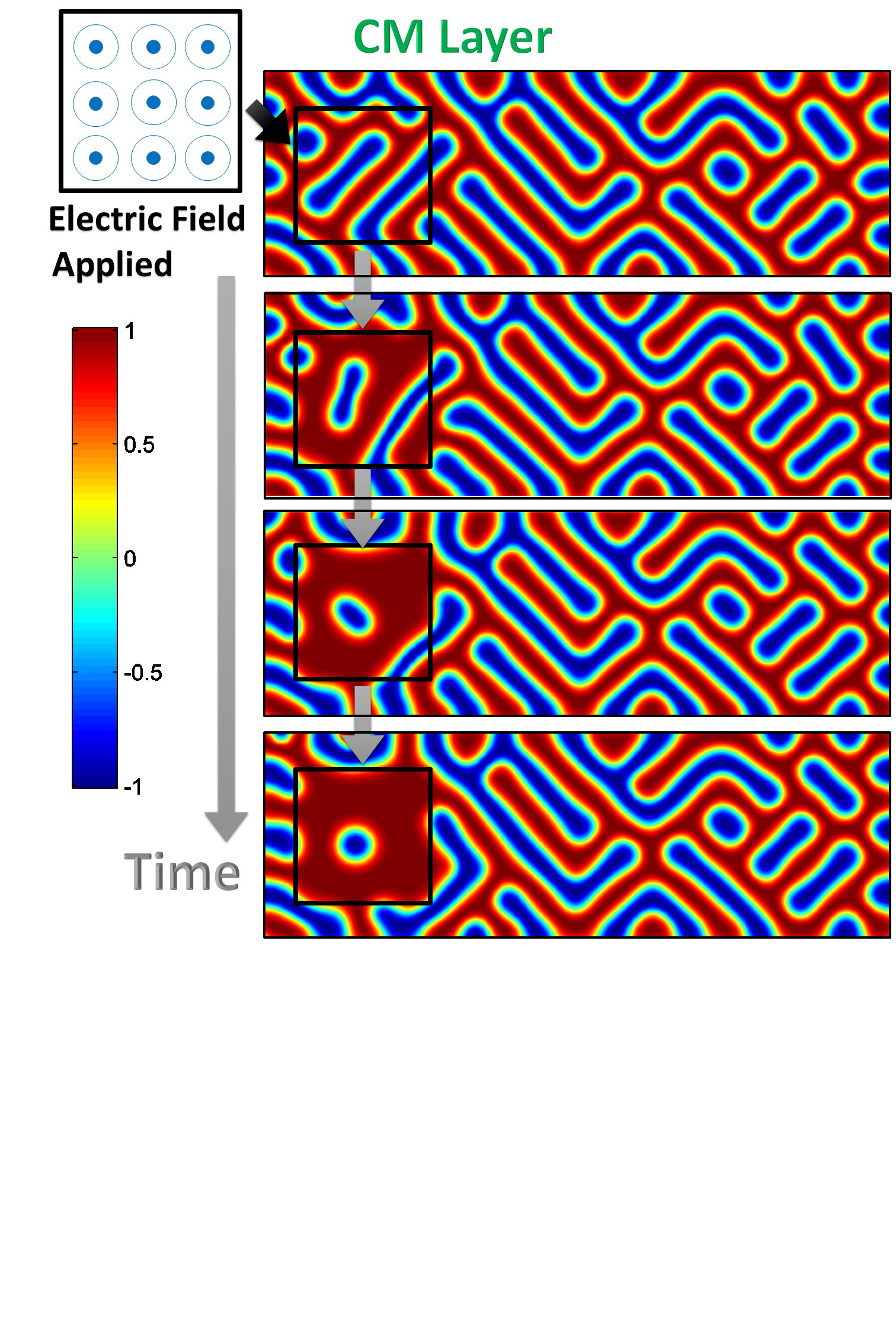}
	\caption{ Sequential snapshots present the generation of a Skyrmion in the bulk of CM layer, as the localized electric field is statically applied at the initial position. The color scale represents the magnitude of the local \textit{z} componential magnetization. }
	\label{Fig.2}
\end{figure}

\begin{figure}
	\includegraphics[width = 250px, trim = 0 140 180 0, clip]{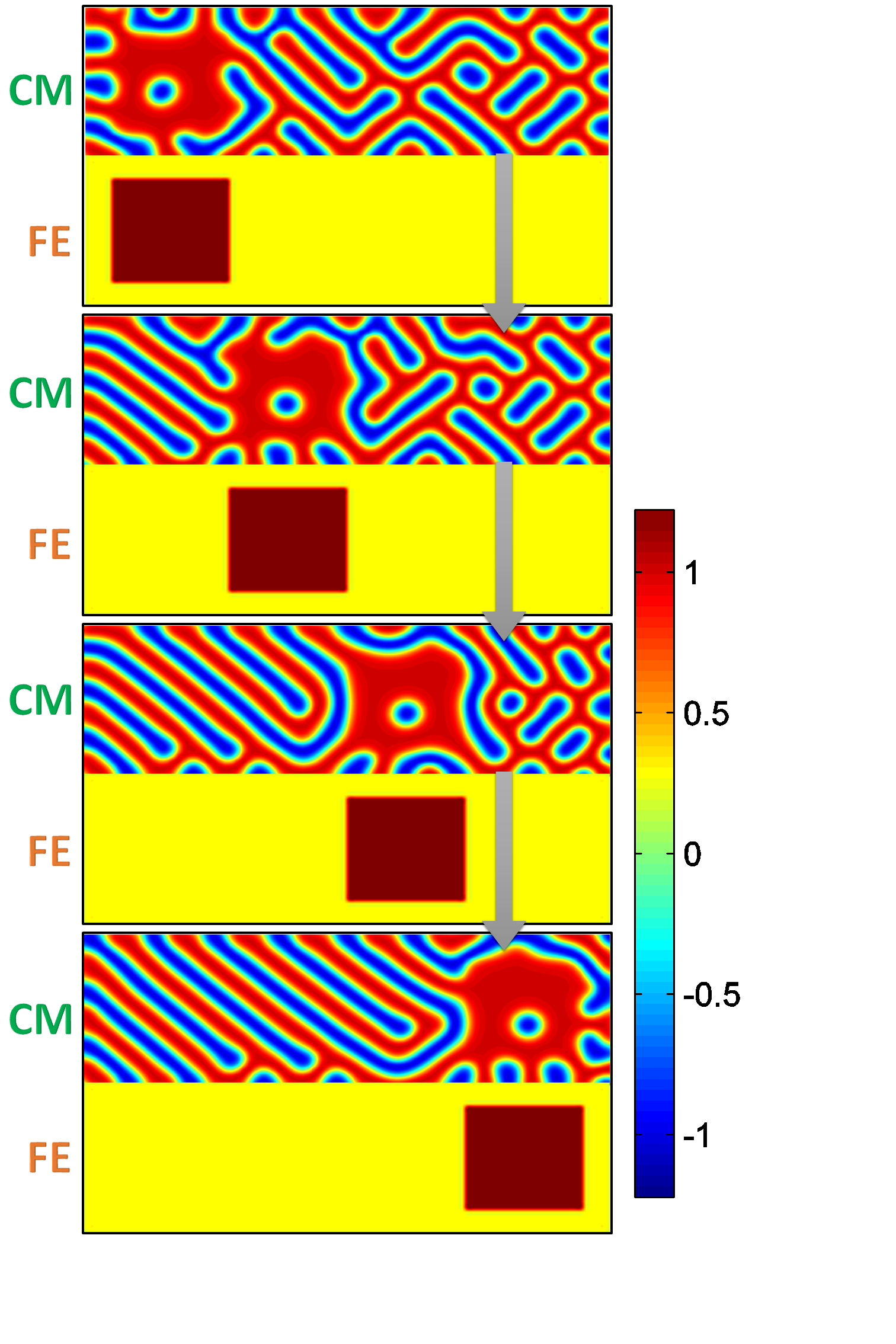}
	\caption{ Propagating a Skyrmion by moving the electric field source with a velocity of $v^* = 0.02$. The color scale represents the magnitude of the \textit{z} component. }
	\label{Fig.3}
\end{figure}

The movement of the field source is externally controllable. We therefore explore the effects of higher traveling speed on the Skyrmion transport. Two results of the Skyrmion transport with different velocities are present in \textbf{FIG.~\ref{Fig.4}} (\textbf{Movie 4} in Supplemental Material \cite{Supplementary.8}). The first case with $v^* = 0.05$ is shown in \textbf{FIG.~\ref{Fig.4}(a)}. The Skyrmion barely struggles to follow the motion of field source during the propagation process. Eventually, the system becomes much more complicated, because another two Skyrmions are formed from edge-merons to complement the energy contribution. In \textbf{FIG.~\ref{Fig.4}(b)}, we double the velocity to $v^* = 0.1$, and note the Skyrmion been lost immediately. Furthermore, the Skyrmion Hall effect acts in the high speed operation, and the transverse motion of Skyrmion transport may result in its annihilation at boundaries.

\begin{figure}
	\includegraphics[width = 250px, trim = 0 720 0 0, clip]{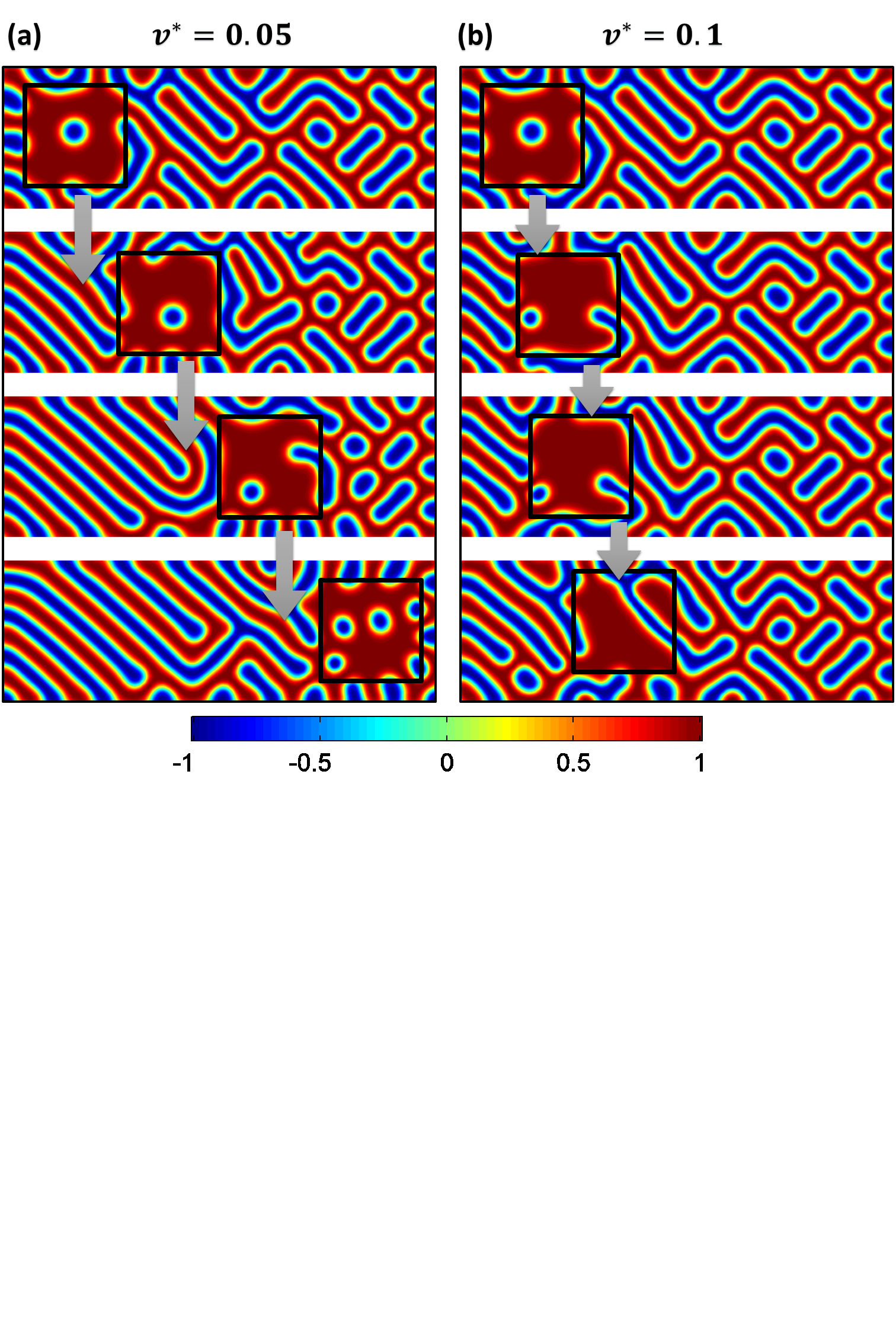}
	\caption{ High velocities effects in Skyrmion transport for (a) $v^* = 0.05$ and (b) $v^* = 0.1$. The color scale represents the magnitude of the \textit{z} component magnetization. }
	\label{Fig.4}
\end{figure}

\textit{Conclusion.} - To summarize, we have investigated a novel mechanical method to control magnetic Bloch Skyrmions by moving a electric field source parallel to the composite CM/FE bilayer system. Skyrmions are supported by the electric polarization through the converse ME effect. The results demonstrate that the Skyrmion is moved collinearly with the field source at a slow speed. But high traveling speeds may break the stability of Skyrmion transport, and annihilate the Skyrmions at edges.

\begin{acknowledgments}
	Z.W. gratefully acknowledges Wang Yuhua, Zhao Bingjin, Zhao Wenxia and Wang Feng for support.
\end{acknowledgments}

\bibliography{Bibliography}

\begin{thebibliography}{29}%
\makeatletter
\providecommand \@ifxundefined [1]{%
 \@ifx{#1\undefined}
}%
\providecommand \@ifnum [1]{%
 \ifnum #1\expandafter \@firstoftwo
 \else \expandafter \@secondoftwo
 \fi
}%
\providecommand \@ifx [1]{%
 \ifx #1\expandafter \@firstoftwo
 \else \expandafter \@secondoftwo
 \fi
}%
\providecommand \natexlab [1]{#1}%
\providecommand \enquote  [1]{``#1''}%
\providecommand \bibnamefont  [1]{#1}%
\providecommand \bibfnamefont [1]{#1}%
\providecommand \citenamefont [1]{#1}%
\providecommand \href@noop [0]{\@secondoftwo}%
\providecommand \href [0]{\begingroup \@sanitize@url \@href}%
\providecommand \@href[1]{\@@startlink{#1}\@@href}%
\providecommand \@@href[1]{\endgroup#1\@@endlink}%
\providecommand \@sanitize@url [0]{\catcode `\\12\catcode `\$12\catcode
  `\&12\catcode `\#12\catcode `\^12\catcode `\_12\catcode `\%12\relax}%
\providecommand \@@startlink[1]{}%
\providecommand \@@endlink[0]{}%
\providecommand \url  [0]{\begingroup\@sanitize@url \@url }%
\providecommand \@url [1]{\endgroup\@href {#1}{\urlprefix }}%
\providecommand \urlprefix  [0]{URL }%
\providecommand \Eprint [0]{\href }%
\providecommand \doibase [0]{http://dx.doi.org/}%
\providecommand \selectlanguage [0]{\@gobble}%
\providecommand \bibinfo  [0]{\@secondoftwo}%
\providecommand \bibfield  [0]{\@secondoftwo}%
\providecommand \translation [1]{[#1]}%
\providecommand \BibitemOpen [0]{}%
\providecommand \bibitemStop [0]{}%
\providecommand \bibitemNoStop [0]{.\EOS\space}%
\providecommand \EOS [0]{\spacefactor3000\relax}%
\providecommand \BibitemShut  [1]{\csname bibitem#1\endcsname}%
\let\auto@bib@innerbib\@empty
\bibitem [{\citenamefont {R{\"o}{\ss}ler}\ \emph {et~al.}(2006)\citenamefont
  {R{\"o}{\ss}ler}, \citenamefont {Bogdanov},\ and\ \citenamefont
  {Pfleiderer}}]{N.442.797}%
  \BibitemOpen
  \bibfield  {author} {\bibinfo {author} {\bibfnamefont {U.}~\bibnamefont
  {R{\"o}{\ss}ler}}, \bibinfo {author} {\bibfnamefont {A.}~\bibnamefont
  {Bogdanov}}, \ and\ \bibinfo {author} {\bibfnamefont {C.}~\bibnamefont
  {Pfleiderer}},\ }\href {\doibase 10.1038/nature05056} {\bibfield  {journal}
  {\bibinfo  {journal} {Nature}\ }\textbf {\bibinfo {volume} {442}},\ \bibinfo
  {pages} {797} (\bibinfo {year} {2006})}\BibitemShut {NoStop}%
\bibitem [{\citenamefont {Nagaosa}\ and\ \citenamefont
  {Tokura}(2013)}]{NN.8.899}%
  \BibitemOpen
  \bibfield  {author} {\bibinfo {author} {\bibfnamefont {N.}~\bibnamefont
  {Nagaosa}}\ and\ \bibinfo {author} {\bibfnamefont {Y.}~\bibnamefont
  {Tokura}},\ }\href {\doibase 10.1038/nnano.2013.243} {\bibfield  {journal}
  {\bibinfo  {journal} {Nat. Nanotech.}\ }\textbf {\bibinfo {volume} {8}},\
  \bibinfo {pages} {899} (\bibinfo {year} {2013})}\BibitemShut {NoStop}%
\bibitem [{\citenamefont {M{\"u}hlbauer}\ \emph {et~al.}(2009)\citenamefont
  {M{\"u}hlbauer}, \citenamefont {Binz}, \citenamefont {Jonietz}, \citenamefont
  {Pfleiderer}, \citenamefont {Rosch}, \citenamefont {Neubauer}, \citenamefont
  {Georgii},\ and\ \citenamefont {B{\"o}ni}}]{S.323.915}%
  \BibitemOpen
  \bibfield  {author} {\bibinfo {author} {\bibfnamefont {S.}~\bibnamefont
  {M{\"u}hlbauer}}, \bibinfo {author} {\bibfnamefont {B.}~\bibnamefont {Binz}},
  \bibinfo {author} {\bibfnamefont {F.}~\bibnamefont {Jonietz}}, \bibinfo
  {author} {\bibfnamefont {C.}~\bibnamefont {Pfleiderer}}, \bibinfo {author}
  {\bibfnamefont {A.}~\bibnamefont {Rosch}}, \bibinfo {author} {\bibfnamefont
  {A.}~\bibnamefont {Neubauer}}, \bibinfo {author} {\bibfnamefont
  {R.}~\bibnamefont {Georgii}}, \ and\ \bibinfo {author} {\bibfnamefont
  {P.}~\bibnamefont {B{\"o}ni}},\ }\href {\doibase 10.1126/science.1166767}
  {\bibfield  {journal} {\bibinfo  {journal} {Science}\ }\textbf {\bibinfo
  {volume} {323}},\ \bibinfo {pages} {915} (\bibinfo {year}
  {2009})}\BibitemShut {NoStop}%
\bibitem [{\citenamefont {Heinze}\ \emph {et~al.}(2011)\citenamefont {Heinze},
  \citenamefont {Von~Bergmann}, \citenamefont {Menzel}, \citenamefont {Brede},
  \citenamefont {Kubetzka}, \citenamefont {Wiesendanger}, \citenamefont
  {Bihlmayer},\ and\ \citenamefont {Bl{\"u}gel}}]{NP.7.713}%
  \BibitemOpen
  \bibfield  {author} {\bibinfo {author} {\bibfnamefont {S.}~\bibnamefont
  {Heinze}}, \bibinfo {author} {\bibfnamefont {K.}~\bibnamefont
  {Von~Bergmann}}, \bibinfo {author} {\bibfnamefont {M.}~\bibnamefont
  {Menzel}}, \bibinfo {author} {\bibfnamefont {J.}~\bibnamefont {Brede}},
  \bibinfo {author} {\bibfnamefont {A.}~\bibnamefont {Kubetzka}}, \bibinfo
  {author} {\bibfnamefont {R.}~\bibnamefont {Wiesendanger}}, \bibinfo {author}
  {\bibfnamefont {G.}~\bibnamefont {Bihlmayer}}, \ and\ \bibinfo {author}
  {\bibfnamefont {S.}~\bibnamefont {Bl{\"u}gel}},\ }\href {\doibase
  10.1038/nphys2045} {\bibfield  {journal} {\bibinfo  {journal} {Nat. Phys.}\
  }\textbf {\bibinfo {volume} {7}},\ \bibinfo {pages} {713} (\bibinfo {year}
  {2011})}\BibitemShut {NoStop}%
\bibitem [{\citenamefont {Yu}\ \emph {et~al.}(2011)\citenamefont {Yu},
  \citenamefont {Kanazawa}, \citenamefont {Onose}, \citenamefont {Kimoto},
  \citenamefont {Zhang}, \citenamefont {Ishiwata}, \citenamefont {Matsui},\
  and\ \citenamefont {Tokura}}]{NM.10.106}%
  \BibitemOpen
  \bibfield  {author} {\bibinfo {author} {\bibfnamefont {X.~Z.}\ \bibnamefont
  {Yu}}, \bibinfo {author} {\bibfnamefont {N.}~\bibnamefont {Kanazawa}},
  \bibinfo {author} {\bibfnamefont {Y.}~\bibnamefont {Onose}}, \bibinfo
  {author} {\bibfnamefont {K.}~\bibnamefont {Kimoto}}, \bibinfo {author}
  {\bibfnamefont {W.~Z.}\ \bibnamefont {Zhang}}, \bibinfo {author}
  {\bibfnamefont {S.}~\bibnamefont {Ishiwata}}, \bibinfo {author}
  {\bibfnamefont {Y.}~\bibnamefont {Matsui}}, \ and\ \bibinfo {author}
  {\bibfnamefont {Y.}~\bibnamefont {Tokura}},\ }\href {\doibase
  10.1038/nmat2916} {\bibfield  {journal} {\bibinfo  {journal} {Nat. Mater.}\
  }\textbf {\bibinfo {volume} {10}},\ \bibinfo {pages} {106} (\bibinfo {year}
  {2011})}\BibitemShut {NoStop}%
\bibitem [{\citenamefont {Shibata}\ \emph {et~al.}(2013)\citenamefont
  {Shibata}, \citenamefont {Yu}, \citenamefont {Hara}, \citenamefont
  {Morikawa}, \citenamefont {Kanazawa}, \citenamefont {Kimoto}, \citenamefont
  {Ishiwata}, \citenamefont {Matsui},\ and\ \citenamefont {Tokura}}]{NN.8.723}%
  \BibitemOpen
  \bibfield  {author} {\bibinfo {author} {\bibfnamefont {K.}~\bibnamefont
  {Shibata}}, \bibinfo {author} {\bibfnamefont {X.}~\bibnamefont {Yu}},
  \bibinfo {author} {\bibfnamefont {T.}~\bibnamefont {Hara}}, \bibinfo {author}
  {\bibfnamefont {D.}~\bibnamefont {Morikawa}}, \bibinfo {author}
  {\bibfnamefont {N.}~\bibnamefont {Kanazawa}}, \bibinfo {author}
  {\bibfnamefont {K.}~\bibnamefont {Kimoto}}, \bibinfo {author} {\bibfnamefont
  {S.}~\bibnamefont {Ishiwata}}, \bibinfo {author} {\bibfnamefont
  {Y.}~\bibnamefont {Matsui}}, \ and\ \bibinfo {author} {\bibfnamefont
  {Y.}~\bibnamefont {Tokura}},\ }\href {\doibase 10.1038/nnano.2013.174}
  {\bibfield  {journal} {\bibinfo  {journal} {Nat. Nanotech.}\ }\textbf
  {\bibinfo {volume} {8}},\ \bibinfo {pages} {723} (\bibinfo {year}
  {2013})}\BibitemShut {NoStop}%
\bibitem [{\citenamefont {K\'{e}zsm\'{a}rki}\ \emph {et~al.}(2015)\citenamefont
  {K\'{e}zsm\'{a}rki}, \citenamefont {Bordacs}, \citenamefont {Milde},
  \citenamefont {Neuber}, \citenamefont {Eng}, \citenamefont {White},
  \citenamefont {R\o{}nnow}, \citenamefont {Dewhurst}, \citenamefont
  {Mochizuki}, \citenamefont {Yanai}, \citenamefont {Nakamura}, \citenamefont
  {Ehlers}, \citenamefont {Tsurkan},\ and\ \citenamefont {Loidl}}]{NM.14.1116}%
  \BibitemOpen
  \bibfield  {author} {\bibinfo {author} {\bibfnamefont {I.}~\bibnamefont
  {K\'{e}zsm\'{a}rki}}, \bibinfo {author} {\bibfnamefont {S.}~\bibnamefont
  {Bordacs}}, \bibinfo {author} {\bibfnamefont {P.}~\bibnamefont {Milde}},
  \bibinfo {author} {\bibfnamefont {E.}~\bibnamefont {Neuber}}, \bibinfo
  {author} {\bibfnamefont {L.~M.}\ \bibnamefont {Eng}}, \bibinfo {author}
  {\bibfnamefont {J.~S.}\ \bibnamefont {White}}, \bibinfo {author}
  {\bibfnamefont {H.~M.}\ \bibnamefont {R\o{}nnow}}, \bibinfo {author}
  {\bibfnamefont {C.~D.}\ \bibnamefont {Dewhurst}}, \bibinfo {author}
  {\bibfnamefont {M.}~\bibnamefont {Mochizuki}}, \bibinfo {author}
  {\bibfnamefont {K.}~\bibnamefont {Yanai}}, \bibinfo {author} {\bibfnamefont
  {H.}~\bibnamefont {Nakamura}}, \bibinfo {author} {\bibfnamefont
  {D.}~\bibnamefont {Ehlers}}, \bibinfo {author} {\bibfnamefont
  {V.}~\bibnamefont {Tsurkan}}, \ and\ \bibinfo {author} {\bibfnamefont
  {A.}~\bibnamefont {Loidl}},\ }\href {\doibase 10.1038/nmat4402} {\bibfield
  {journal} {\bibinfo  {journal} {Nat. Mater.}\ }\textbf {\bibinfo {volume}
  {14}},\ \bibinfo {pages} {1116} (\bibinfo {year} {2015})}\BibitemShut
  {NoStop}%
\bibitem [{\citenamefont {Adams}\ \emph {et~al.}(2012)\citenamefont {Adams},
  \citenamefont {Chacon}, \citenamefont {Wagner}, \citenamefont {Bauer},
  \citenamefont {Brandl}, \citenamefont {Pedersen}, \citenamefont {Berger},
  \citenamefont {Lemmens},\ and\ \citenamefont {Pfleiderer}}]{PRL.108.237204}%
  \BibitemOpen
  \bibfield  {author} {\bibinfo {author} {\bibfnamefont {T.}~\bibnamefont
  {Adams}}, \bibinfo {author} {\bibfnamefont {A.}~\bibnamefont {Chacon}},
  \bibinfo {author} {\bibfnamefont {M.}~\bibnamefont {Wagner}}, \bibinfo
  {author} {\bibfnamefont {A.}~\bibnamefont {Bauer}}, \bibinfo {author}
  {\bibfnamefont {G.}~\bibnamefont {Brandl}}, \bibinfo {author} {\bibfnamefont
  {B.}~\bibnamefont {Pedersen}}, \bibinfo {author} {\bibfnamefont
  {H.}~\bibnamefont {Berger}}, \bibinfo {author} {\bibfnamefont
  {P.}~\bibnamefont {Lemmens}}, \ and\ \bibinfo {author} {\bibfnamefont
  {C.}~\bibnamefont {Pfleiderer}},\ }\href {\doibase
  10.1103/PhysRevLett.108.237204} {\bibfield  {journal} {\bibinfo  {journal}
  {Phys. Rev. Lett.}\ }\textbf {\bibinfo {volume} {108}},\ \bibinfo {pages}
  {237204} (\bibinfo {year} {2012})}\BibitemShut {NoStop}%
\bibitem [{\citenamefont {M\"unzer}\ \emph {et~al.}(2010)\citenamefont
  {M\"unzer}, \citenamefont {Neubauer}, \citenamefont {Adams}, \citenamefont
  {M\"uhlbauer}, \citenamefont {Franz}, \citenamefont {Jonietz}, \citenamefont
  {Georgii}, \citenamefont {B\"oni}, \citenamefont {Pedersen}, \citenamefont
  {Schmidt}, \citenamefont {Rosch},\ and\ \citenamefont
  {Pfleiderer}}]{PRB.81.041203}%
  \BibitemOpen
  \bibfield  {author} {\bibinfo {author} {\bibfnamefont {W.}~\bibnamefont
  {M\"unzer}}, \bibinfo {author} {\bibfnamefont {A.}~\bibnamefont {Neubauer}},
  \bibinfo {author} {\bibfnamefont {T.}~\bibnamefont {Adams}}, \bibinfo
  {author} {\bibfnamefont {S.}~\bibnamefont {M\"uhlbauer}}, \bibinfo {author}
  {\bibfnamefont {C.}~\bibnamefont {Franz}}, \bibinfo {author} {\bibfnamefont
  {F.}~\bibnamefont {Jonietz}}, \bibinfo {author} {\bibfnamefont
  {R.}~\bibnamefont {Georgii}}, \bibinfo {author} {\bibfnamefont
  {P.}~\bibnamefont {B\"oni}}, \bibinfo {author} {\bibfnamefont
  {B.}~\bibnamefont {Pedersen}}, \bibinfo {author} {\bibfnamefont
  {M.}~\bibnamefont {Schmidt}}, \bibinfo {author} {\bibfnamefont
  {A.}~\bibnamefont {Rosch}}, \ and\ \bibinfo {author} {\bibfnamefont
  {C.}~\bibnamefont {Pfleiderer}},\ }\href {\doibase
  10.1103/PhysRevB.81.041203} {\bibfield  {journal} {\bibinfo  {journal} {Phys.
  Rev. B}\ }\textbf {\bibinfo {volume} {81}},\ \bibinfo {pages} {041203}
  (\bibinfo {year} {2010})}\BibitemShut {NoStop}%
\bibitem [{\citenamefont {Woo}\ \emph {et~al.}(2016)\citenamefont {Woo},
  \citenamefont {Litzius}, \citenamefont {Kruger}, \citenamefont {Im},
  \citenamefont {Caretta}, \citenamefont {Richter}, \citenamefont {Mann},
  \citenamefont {Krone}, \citenamefont {Reeve}, \citenamefont {Weigand},
  \citenamefont {Agrawal}, \citenamefont {Lemesh}, \citenamefont {Mawass},
  \citenamefont {Fischer}, \citenamefont {Klaui},\ and\ \citenamefont
  {Beach}}]{NM.15.501}%
  \BibitemOpen
  \bibfield  {author} {\bibinfo {author} {\bibfnamefont {S.}~\bibnamefont
  {Woo}}, \bibinfo {author} {\bibfnamefont {K.}~\bibnamefont {Litzius}},
  \bibinfo {author} {\bibfnamefont {B.}~\bibnamefont {Kruger}}, \bibinfo
  {author} {\bibfnamefont {M.-Y.}\ \bibnamefont {Im}}, \bibinfo {author}
  {\bibfnamefont {L.}~\bibnamefont {Caretta}}, \bibinfo {author} {\bibfnamefont
  {K.}~\bibnamefont {Richter}}, \bibinfo {author} {\bibfnamefont
  {M.}~\bibnamefont {Mann}}, \bibinfo {author} {\bibfnamefont {A.}~\bibnamefont
  {Krone}}, \bibinfo {author} {\bibfnamefont {R.~M.}\ \bibnamefont {Reeve}},
  \bibinfo {author} {\bibfnamefont {M.}~\bibnamefont {Weigand}}, \bibinfo
  {author} {\bibfnamefont {P.}~\bibnamefont {Agrawal}}, \bibinfo {author}
  {\bibfnamefont {I.}~\bibnamefont {Lemesh}}, \bibinfo {author} {\bibfnamefont
  {M.-A.}\ \bibnamefont {Mawass}}, \bibinfo {author} {\bibfnamefont
  {P.}~\bibnamefont {Fischer}}, \bibinfo {author} {\bibfnamefont
  {M.}~\bibnamefont {Klaui}}, \ and\ \bibinfo {author} {\bibfnamefont
  {G.~S.~D.}\ \bibnamefont {Beach}},\ }\href {\doibase 10.1038/nmat4593}
  {\bibfield  {journal} {\bibinfo  {journal} {Nat. Mater.}\ }\textbf {\bibinfo
  {volume} {15}},\ \bibinfo {pages} {501} (\bibinfo {year} {2016})}\BibitemShut
  {NoStop}%
\bibitem [{\citenamefont {Fert}\ \emph {et~al.}(2013)\citenamefont {Fert},
  \citenamefont {Cros},\ and\ \citenamefont {Sampaio}}]{NN.8.152}%
  \BibitemOpen
  \bibfield  {author} {\bibinfo {author} {\bibfnamefont {A.}~\bibnamefont
  {Fert}}, \bibinfo {author} {\bibfnamefont {V.}~\bibnamefont {Cros}}, \ and\
  \bibinfo {author} {\bibfnamefont {J.~a.}\ \bibnamefont {Sampaio}},\ }\href
  {\doibase 10.1038/nnano.2013.29} {\bibfield  {journal} {\bibinfo  {journal}
  {Nat. Nanotech.}\ }\textbf {\bibinfo {volume} {8}},\ \bibinfo {pages} {152}
  (\bibinfo {year} {2013})}\BibitemShut {NoStop}%
\bibitem [{\citenamefont {Iwasaki}\ \emph {et~al.}(2013)\citenamefont
  {Iwasaki}, \citenamefont {Nagaosa},\ and\ \citenamefont {Tokura}}]{NN.8.742}%
  \BibitemOpen
  \bibfield  {author} {\bibinfo {author} {\bibfnamefont {J.}~\bibnamefont
  {Iwasaki}}, \bibinfo {author} {\bibfnamefont {N.}~\bibnamefont {Nagaosa}}, \
  and\ \bibinfo {author} {\bibfnamefont {Y.}~\bibnamefont {Tokura}},\ }\href
  {\doibase 10.1038/nnano.2013.176} {\bibfield  {journal} {\bibinfo  {journal}
  {Nat. Nanotech.}\ }\textbf {\bibinfo {volume} {8}},\ \bibinfo {pages} {742}
  (\bibinfo {year} {2013})}\BibitemShut {NoStop}%
\bibitem [{\citenamefont {Wang}\ \emph {et~al.}(2015)\citenamefont {Wang},
  \citenamefont {Beg}, \citenamefont {Zhang}, \citenamefont {Kuch},\ and\
  \citenamefont {Fangohr}}]{PRB.92.020403}%
  \BibitemOpen
  \bibfield  {author} {\bibinfo {author} {\bibfnamefont {W.}~\bibnamefont
  {Wang}}, \bibinfo {author} {\bibfnamefont {M.}~\bibnamefont {Beg}}, \bibinfo
  {author} {\bibfnamefont {B.}~\bibnamefont {Zhang}}, \bibinfo {author}
  {\bibfnamefont {W.}~\bibnamefont {Kuch}}, \ and\ \bibinfo {author}
  {\bibfnamefont {H.}~\bibnamefont {Fangohr}},\ }\href {\doibase
  10.1103/PhysRevB.92.020403} {\bibfield  {journal} {\bibinfo  {journal} {Phys.
  Rev. B}\ }\textbf {\bibinfo {volume} {92}},\ \bibinfo {pages} {020403}
  (\bibinfo {year} {2015})}\BibitemShut {NoStop}%
\bibitem [{\citenamefont {Kong}\ and\ \citenamefont
  {Zang}(2013)}]{PRL.111.067203}%
  \BibitemOpen
  \bibfield  {author} {\bibinfo {author} {\bibfnamefont {L.}~\bibnamefont
  {Kong}}\ and\ \bibinfo {author} {\bibfnamefont {J.}~\bibnamefont {Zang}},\
  }\href {\doibase 10.1103/PhysRevLett.111.067203} {\bibfield  {journal}
  {\bibinfo  {journal} {Phys. Rev. Lett.}\ }\textbf {\bibinfo {volume} {111}},\
  \bibinfo {pages} {067203} (\bibinfo {year} {2013})}\BibitemShut {NoStop}%
\bibitem [{\citenamefont {Lin}\ \emph {et~al.}(2014)\citenamefont {Lin},
  \citenamefont {Batista}, \citenamefont {Reichhardt},\ and\ \citenamefont
  {Saxena}}]{PRL.112.187203}%
  \BibitemOpen
  \bibfield  {author} {\bibinfo {author} {\bibfnamefont {S.-Z.}\ \bibnamefont
  {Lin}}, \bibinfo {author} {\bibfnamefont {C.~D.}\ \bibnamefont {Batista}},
  \bibinfo {author} {\bibfnamefont {C.}~\bibnamefont {Reichhardt}}, \ and\
  \bibinfo {author} {\bibfnamefont {A.}~\bibnamefont {Saxena}},\ }\href
  {\doibase 10.1103/PhysRevLett.112.187203} {\bibfield  {journal} {\bibinfo
  {journal} {Phys. Rev. Lett.}\ }\textbf {\bibinfo {volume} {112}},\ \bibinfo
  {pages} {187203} (\bibinfo {year} {2014})}\BibitemShut {NoStop}%
\bibitem [{\citenamefont {Sch\"utte}\ and\ \citenamefont
  {Garst}(2014)}]{PRB.90.094423}%
  \BibitemOpen
  \bibfield  {author} {\bibinfo {author} {\bibfnamefont {C.}~\bibnamefont
  {Sch\"utte}}\ and\ \bibinfo {author} {\bibfnamefont {M.}~\bibnamefont
  {Garst}},\ }\href {\doibase 10.1103/PhysRevB.90.094423} {\bibfield  {journal}
  {\bibinfo  {journal} {Phys. Rev. B}\ }\textbf {\bibinfo {volume} {90}},\
  \bibinfo {pages} {094423} (\bibinfo {year} {2014})}\BibitemShut {NoStop}%
\bibitem [{\citenamefont {Wang}\ and\ \citenamefont
  {Grimson}(2016{\natexlab{a}})}]{PRB.94.014311}%
  \BibitemOpen
  \bibfield  {author} {\bibinfo {author} {\bibfnamefont {Z.}~\bibnamefont
  {Wang}}\ and\ \bibinfo {author} {\bibfnamefont {M.~J.}\ \bibnamefont
  {Grimson}},\ }\href {\doibase 10.1103/PhysRevB.94.014311} {\bibfield
  {journal} {\bibinfo  {journal} {Phys. Rev. B}\ }\textbf {\bibinfo {volume}
  {94}},\ \bibinfo {pages} {014311} (\bibinfo {year}
  {2016}{\natexlab{a}})}\BibitemShut {NoStop}%
\bibitem [{Sup()}]{Supplementary.8}%
  \BibitemOpen
  \href {\doibase 10.13140/RG.2.2.26175.51360} {\bibfield  {journal} {\bibinfo
  {journal} {See Supplemental Material at}\ }10.13140/RG.2.2.26175.51360},\
  \bibinfo {note} {for animations about dynamical results}\BibitemShut
  {NoStop}%
\bibitem [{\citenamefont {White}\ \emph {et~al.}(2014)\citenamefont {White},
  \citenamefont {Pr\ifmmode~\check{s}\else \v{s}\fi{}a}, \citenamefont {Huang},
  \citenamefont {Omrani}, \citenamefont {\ifmmode \check{Z}\else
  \v{Z}\fi{}ivkovi\ifmmode~\acute{c}\else \'{c}\fi{}}, \citenamefont
  {Bartkowiak}, \citenamefont {Berger}, \citenamefont {Magrez}, \citenamefont
  {Gavilano}, \citenamefont {Nagy}, \citenamefont {Zang},\ and\ \citenamefont
  {R\o{}nnow}}]{PRL.113.107203}%
  \BibitemOpen
  \bibfield  {author} {\bibinfo {author} {\bibfnamefont {J.~S.}\ \bibnamefont
  {White}}, \bibinfo {author} {\bibfnamefont {K.}~\bibnamefont
  {Pr\ifmmode~\check{s}\else \v{s}\fi{}a}}, \bibinfo {author} {\bibfnamefont
  {P.}~\bibnamefont {Huang}}, \bibinfo {author} {\bibfnamefont {A.~A.}\
  \bibnamefont {Omrani}}, \bibinfo {author} {\bibfnamefont {I.}~\bibnamefont
  {\ifmmode \check{Z}\else \v{Z}\fi{}ivkovi\ifmmode~\acute{c}\else
  \'{c}\fi{}}}, \bibinfo {author} {\bibfnamefont {M.}~\bibnamefont
  {Bartkowiak}}, \bibinfo {author} {\bibfnamefont {H.}~\bibnamefont {Berger}},
  \bibinfo {author} {\bibfnamefont {A.}~\bibnamefont {Magrez}}, \bibinfo
  {author} {\bibfnamefont {J.~L.}\ \bibnamefont {Gavilano}}, \bibinfo {author}
  {\bibfnamefont {G.}~\bibnamefont {Nagy}}, \bibinfo {author} {\bibfnamefont
  {J.}~\bibnamefont {Zang}}, \ and\ \bibinfo {author} {\bibfnamefont {H.~M.}\
  \bibnamefont {R\o{}nnow}},\ }\href {\doibase 10.1103/PhysRevLett.113.107203}
  {\bibfield  {journal} {\bibinfo  {journal} {Phys. Rev. Lett.}\ }\textbf
  {\bibinfo {volume} {113}},\ \bibinfo {pages} {107203} (\bibinfo {year}
  {2014})}\BibitemShut {NoStop}%
\bibitem [{\citenamefont {Wang}\ and\ \citenamefont
  {Grimson}(2015)}]{JAP.118.124109}%
  \BibitemOpen
  \bibfield  {author} {\bibinfo {author} {\bibfnamefont {Z.}~\bibnamefont
  {Wang}}\ and\ \bibinfo {author} {\bibfnamefont {M.~J.}\ \bibnamefont
  {Grimson}},\ }\href {\doibase 10.1063/1.4931895} {\bibfield  {journal}
  {\bibinfo  {journal} {J. Appl. Phys.}\ }\textbf {\bibinfo {volume} {118}},\
  \bibinfo {pages} {124109} (\bibinfo {year} {2015})}\BibitemShut {NoStop}%
\bibitem [{\citenamefont {Kittel}(2005)}]{C.Kittel}%
  \BibitemOpen
  \bibfield  {author} {\bibinfo {author} {\bibfnamefont {C.}~\bibnamefont
  {Kittel}},\ }\href@noop {} {\emph {\bibinfo {title} {Introduction to Solid
  State Physics}}}\ (\bibinfo  {publisher} {Wiley, Hoboken, NJ},\ \bibinfo
  {year} {2005})\ Chap.~\bibinfo {chapter} {16}\BibitemShut {NoStop}%
\bibitem [{\citenamefont {Wang}\ and\ \citenamefont
  {Grimson}(2016{\natexlab{b}})}]{JAP.119.124105}%
  \BibitemOpen
  \bibfield  {author} {\bibinfo {author} {\bibfnamefont {Z.}~\bibnamefont
  {Wang}}\ and\ \bibinfo {author} {\bibfnamefont {M.~J.}\ \bibnamefont
  {Grimson}},\ }\href {\doibase 10.1063/1.4944604} {\bibfield  {journal}
  {\bibinfo  {journal} {J. Appl. Phys.}\ }\textbf {\bibinfo {volume} {119}},\
  \bibinfo {pages} {124105} (\bibinfo {year} {2016}{\natexlab{b}})}\BibitemShut
  {NoStop}%
\bibitem [{\citenamefont {de~Gennes}(1963)}]{SSC.1.132}%
  \BibitemOpen
  \bibfield  {author} {\bibinfo {author} {\bibfnamefont {P.~G.}\ \bibnamefont
  {de~Gennes}},\ }\href {\doibase 10.1016/0038-1098(63)90212-6} {\bibfield
  {journal} {\bibinfo  {journal} {Solid State Commun.}\ }\textbf {\bibinfo
  {volume} {1}},\ \bibinfo {pages} {132} (\bibinfo {year} {1963})}\BibitemShut
  {NoStop}%
\bibitem [{\citenamefont {Xing}\ \emph {et~al.}(2016)\citenamefont {Xing},
  \citenamefont {Pong},\ and\ \citenamefont {Zhou}}]{JAP.120.203903}%
  \BibitemOpen
  \bibfield  {author} {\bibinfo {author} {\bibfnamefont {X.}~\bibnamefont
  {Xing}}, \bibinfo {author} {\bibfnamefont {P.~W.~T.}\ \bibnamefont {Pong}}, \
  and\ \bibinfo {author} {\bibfnamefont {Y.}~\bibnamefont {Zhou}},\ }\href
  {\doibase http://dx.doi.org/10.1063/1.4968574} {\bibfield  {journal}
  {\bibinfo  {journal} {J. Appl. Phys.}\ }\textbf {\bibinfo {volume} {120}},\
  \bibinfo {pages} {203903} (\bibinfo {year} {2016})}\BibitemShut {NoStop}%
\bibitem [{\citenamefont {Nan}\ \emph {et~al.}(2008)\citenamefont {Nan},
  \citenamefont {Bichurin}, \citenamefont {Dong}, \citenamefont {Viehland},\
  and\ \citenamefont {Srinivasan}}]{JAP.103.031101}%
  \BibitemOpen
  \bibfield  {author} {\bibinfo {author} {\bibfnamefont {C.-W.}\ \bibnamefont
  {Nan}}, \bibinfo {author} {\bibfnamefont {M.~I.}\ \bibnamefont {Bichurin}},
  \bibinfo {author} {\bibfnamefont {S.}~\bibnamefont {Dong}}, \bibinfo {author}
  {\bibfnamefont {D.}~\bibnamefont {Viehland}}, \ and\ \bibinfo {author}
  {\bibfnamefont {G.}~\bibnamefont {Srinivasan}},\ }\href {\doibase
  10.1063/1.2836410} {\bibfield  {journal} {\bibinfo  {journal} {J. Appl.
  Phys.}\ }\textbf {\bibinfo {volume} {103}},\ \bibinfo {pages} {031101}
  (\bibinfo {year} {2008})}\BibitemShut {NoStop}%
\bibitem [{\citenamefont {Chotorlishvili}\ \emph {et~al.}(2015)\citenamefont
  {Chotorlishvili}, \citenamefont {Etesami}, \citenamefont {Berakdar},
  \citenamefont {Khomeriki},\ and\ \citenamefont {Ren}}]{PRB.92.134424}%
  \BibitemOpen
  \bibfield  {author} {\bibinfo {author} {\bibfnamefont {L.}~\bibnamefont
  {Chotorlishvili}}, \bibinfo {author} {\bibfnamefont {S.~R.}\ \bibnamefont
  {Etesami}}, \bibinfo {author} {\bibfnamefont {J.}~\bibnamefont {Berakdar}},
  \bibinfo {author} {\bibfnamefont {R.}~\bibnamefont {Khomeriki}}, \ and\
  \bibinfo {author} {\bibfnamefont {J.}~\bibnamefont {Ren}},\ }\href {\doibase
  10.1103/PhysRevB.92.134424} {\bibfield  {journal} {\bibinfo  {journal} {Phys.
  Rev. B}\ }\textbf {\bibinfo {volume} {92}},\ \bibinfo {pages} {134424}
  (\bibinfo {year} {2015})}\BibitemShut {NoStop}%
\bibitem [{\citenamefont {Th\"ole}\ \emph {et~al.}(2016)\citenamefont
  {Th\"ole}, \citenamefont {Fechner},\ and\ \citenamefont
  {Spaldin}}]{PRB.93.195167}%
  \BibitemOpen
  \bibfield  {author} {\bibinfo {author} {\bibfnamefont {F.}~\bibnamefont
  {Th\"ole}}, \bibinfo {author} {\bibfnamefont {M.}~\bibnamefont {Fechner}}, \
  and\ \bibinfo {author} {\bibfnamefont {N.~A.}\ \bibnamefont {Spaldin}},\
  }\href {\doibase 10.1103/PhysRevB.93.195167} {\bibfield  {journal} {\bibinfo
  {journal} {Phys. Rev. B}\ }\textbf {\bibinfo {volume} {93}},\ \bibinfo
  {pages} {195167} (\bibinfo {year} {2016})}\BibitemShut {NoStop}%
\bibitem [{\citenamefont {Li}\ \emph {et~al.}(2015)\citenamefont {Li},
  \citenamefont {Xu}, \citenamefont {Hu}, \citenamefont {Suter}, \citenamefont
  {Schreiber}, \citenamefont {Ramuhalli}, \citenamefont {Johnson},\ and\
  \citenamefont {McCloy}}]{JPD.48.305001}%
  \BibitemOpen
  \bibfield  {author} {\bibinfo {author} {\bibfnamefont {Y.}~\bibnamefont
  {Li}}, \bibinfo {author} {\bibfnamefont {K.}~\bibnamefont {Xu}}, \bibinfo
  {author} {\bibfnamefont {S.}~\bibnamefont {Hu}}, \bibinfo {author}
  {\bibfnamefont {J.}~\bibnamefont {Suter}}, \bibinfo {author} {\bibfnamefont
  {D.~K.}\ \bibnamefont {Schreiber}}, \bibinfo {author} {\bibfnamefont
  {P.}~\bibnamefont {Ramuhalli}}, \bibinfo {author} {\bibfnamefont {B.~R.}\
  \bibnamefont {Johnson}}, \ and\ \bibinfo {author} {\bibfnamefont
  {J.}~\bibnamefont {McCloy}},\ }\href {\doibase
  10.1088/0022-3727/48/30/305001} {\bibfield  {journal} {\bibinfo  {journal}
  {J. Phys. D Appl. Phys.}\ }\textbf {\bibinfo {volume} {48}},\ \bibinfo
  {pages} {305001} (\bibinfo {year} {2015})}\BibitemShut {NoStop}%
\bibitem [{\citenamefont {Jiang}\ \emph {et~al.}(2016)\citenamefont {Jiang},
  \citenamefont {Zhang}, \citenamefont {Yu}, \citenamefont {Zhang},
  \citenamefont {Jungfleisch}, \citenamefont {Pearson}, \citenamefont
  {Heinonen}, \citenamefont {Wang}, \citenamefont {Zhou}, \citenamefont
  {Hoffmann},\ and\ \citenamefont {te~Velthuis}}]{NP.2016}%
  \BibitemOpen
  \bibfield  {author} {\bibinfo {author} {\bibfnamefont {W.}~\bibnamefont
  {Jiang}}, \bibinfo {author} {\bibfnamefont {X.}~\bibnamefont {Zhang}},
  \bibinfo {author} {\bibfnamefont {G.}~\bibnamefont {Yu}}, \bibinfo {author}
  {\bibfnamefont {W.}~\bibnamefont {Zhang}}, \bibinfo {author} {\bibfnamefont
  {M.~B.}\ \bibnamefont {Jungfleisch}}, \bibinfo {author} {\bibfnamefont
  {J.~E.}\ \bibnamefont {Pearson}}, \bibinfo {author} {\bibfnamefont
  {O.}~\bibnamefont {Heinonen}}, \bibinfo {author} {\bibfnamefont {K.~L.}\
  \bibnamefont {Wang}}, \bibinfo {author} {\bibfnamefont {Y.}~\bibnamefont
  {Zhou}}, \bibinfo {author} {\bibfnamefont {A.}~\bibnamefont {Hoffmann}}, \
  and\ \bibinfo {author} {\bibfnamefont {S.~G.~E.}\ \bibnamefont
  {te~Velthuis}},\ }\href {\doibase 10.1038/nphys3883} {\bibfield  {journal}
  {\bibinfo  {journal} {Nat. Phys.}\ } (\bibinfo {year} {2016}),\
  10.1038/nphys3883}\BibitemShut {NoStop}%
\end{thebibliography}%

\end{document}